\documentclass[twocolumn]{aastex63}

\begin{document}

\title{A Search for Millilensing Gamma-Ray Bursts in the Observations of \textit{Fermi} GBM}

\correspondingauthor{He Gao}
\email{gaohe@bnu.edu.cn}
\correspondingauthor{Lin Lin}
\email{llin@bnu.edu.cn}

\author[0000-0003-1859-2780]{Shi-Jie Lin}
\affiliation{Department of Astronomy , Beijing Normal University, Beijing 100875, China}

\author{An Li}
\affiliation{Department of Astronomy , Beijing Normal University, Beijing 100875, China}

\author{He Gao}
\affiliation{Department of Astronomy , Beijing Normal University, Beijing 100875, China}

\author{Lin Lin}
\affiliation{Department of Astronomy , Beijing Normal University, Beijing 100875, China}

\author[0000-0003-4111-5958]{Bin-Bin Zhang}
\affiliation{School of Astronomy and Space Science, Nanjing
University, Nanjing 210093, China}
\affiliation{Key Laboratory of Modern Astronomy and Astrophysics (Nanjing University), Ministry of Education, Nanjing 210093, China}

\author{Zi-Ke Liu}
\affiliation{Key Laboratory of Modern Astronomy and Astrophysics (Nanjing University), Ministry of Education, Nanjing 210093, China}

\author[0000-0003-4011-2608]{Jin-Hang Zou}
\affiliation{College of Physics, Hebei Normal University, Shijiazhuang 050024, China}
\affiliation{School of Astronomy and Space Science, Nanjing
University, Nanjing 210093, China}

\author{Zhao Zhang}
\affiliation{Key Laboratory of Modern Astronomy and Astrophysics (Nanjing University), Ministry of Education, Nanjing 210093, China}

\author{Huan Zhou}
\affiliation{School of Physics and Astronomy, Sun Yat-sen University, Zhuhai, 519082, China}

\author{Zheng-Xiang Li}
\affiliation{Department of Astronomy , Beijing Normal University, Beijing 100875, China}

\author{Lin Lan}
\affiliation{Department of Astronomy , Beijing Normal University, Beijing 100875, China}

\begin{abstract}

Millilensing of Gamma-Ray Bursts (GRBs) is expected to manifest as multiple emission episodes in a single triggered GRB with similar light-curve patterns and similar spectrum properties. Identifying such lensed GRBs could help improve constraints on the abundance of compact dark matter. Here we present a systemic search for millilensing among 3000 GRBs observed by the \textit{Fermi} GBM up to 2021 April. Eventually we find 4 interesting candidates by performing auto-correlation test, hardness test, and time-integrated/resolved spectrum test. GRB 081126A and GRB 090717A are ranked as the first class candidate based on their excellent performance both in temporal and spectrum analysis. GRB 081122A and GRB 110517B are ranked as the second class candidates (suspected candidates), mainly because their two emission episodes show clear deviations in part of the time-resolved spectrum or in the time-integrated spectrum. Considering a point mass model for the gravitational lens, our results suggest that the density parameter of lens objects with mass $M_{\rm L}\sim10^{6} M_{\odot}$ is larger than $1.5\times10^{-3}$.  
\end{abstract}


\section{Introduction} \label{sec:intro}

In general relativity, light from very distant sources would be deflected by intervening masses, which is called gravitational lensing effect. Depending on the positions of the source, lens and observer, and the mass and shape of the lens, gravitational lensing can manifest itself by making one source produce multiple images, or altering the source image through shearing and convergence \citep{Schneider1992}. 
Due to the rapid progress in time domain surveys, many searches have been carried out for gravitational lensing of explosive transients, because such systems have been widely proposed as promising cosmological and astrophysical probes~\cite[][for a review]{Oguri2019}. For instance, transients strongly lensed by intervening galaxies could be used to improve constraints on cosmological parameters such as the Hubble constant $H_0$~\citep{2017NatCo...8.1148L,2018NatCo...9.3833L}, difference of time delays between multiple images among different particles, energy, or messengers, could be used for testing fundamental physics from the propagation speed~\citep{2009MNRAS.396..946B,2017PhRvL.118i1102F,2017PhRvL.118i1101C}, statistical search results of lensed transients with different time scales could be applied to derive constraints on the abundance of compact dark matter in different mass ranges~\citep{2016PhRvL.117i1301M,2018PhRvD..98l3523J,2020MNRAS.495.2002L,Zhou2021b}.

As one of the most violent explosions in the Universe, Gamma-Ray Bursts (GRBs) are bright enough to be detected in high-redshift range up to at least $\rm z\sim10$ \citep{Tanvir2009,Salvaterra2009}, so they have long been proposed as the most promising transients for searching for gravitational lensing effect \citep{Paczynski1987,Mao1992,Li2014}. Thanks to the successful operation of several dedicated detectors, e.g., the Burst And Transient Source Experiment (BATSE) on Compton Gamma Ray Observatory (CGRO) \citep{Meegan1992}, the Burst Alert Telescope (BAT) on the Neil Gehrels Swift Observatory \citep{Gehrels2004,Barthelmy2005} and the Gamma-Ray Burst Monitor (GBM) on the Fermi Observatory \citep{Meegan2009}, $\sim10^4$ GRBs have been detected. Two kinds of searches for lensed GRBs have been widely carried out: 
\begin{itemize}
    \item searching for independently triggered GRB pairs with similar light curve and spectrum, but with different flux and a small positional offset \citep{Nemiroff1994,Veres2009,Davidson2011,Li2014,Hurley2019,Ahlgren2020} \footnote{Recently, \cite{Chen2021} suggests that the multi-band afterglow data should also be considered in the search for lensed GRB events.}. This is usually called macrolensing event, for which the lens candidates are galaxies or clusters of galaxies (with mass $>10^7 \rm M_{\odot}$).
    \item searching for two emission episodes in a single triggered GRB with similar light-curve patterns and similar spectrum properties. This is usually called millilensing event, for which the lens candidates could be an intermediate-mass black hole \citep{Paynter2021}, compact dark matter \citep{Ji2018,Nemiroff1993}, star cluster or population III star \citep{Hirose2006}, with mass $10^4~\sim~10^7 \rm M_{\odot}$. 
\end{itemize}

Up to now, all searches for macrolensing events have yielded null results. On the other hand, however, several candidates of millilensing events have been proposed. \cite{Paynter2021} claims the first convincing evidence for millilensing events in the light curve of BATSE GRB 950830 and thus claims the existence of intermediate-mass black holes\footnote{Note that other searches based on the data of BATSE find none well-proved candidates \citep{Nemiroff1993,Nemiroff1994,Nemiroff2001,Ougolnikov2003,Hirose2006,Ji2018}.}. Later on, an increase of millilensing GRB candidates from the data of Fermi/GBM have been claimed \citep{Kalantari2021,Wang2021,Yang2021,veres2021fermigbm}. For instance, \cite{Wang2021} and \cite{Yang2021} made independent analysis for GRB 200716C and argued it showing millilensing signatures. \cite{veres2021fermigbm} made an exhaustive temporal and spectral analysis to claim that GRB 210812A shows strong evidence in favor of the millilensing effects. 

Inspired by the claims of these individual cases, it is essential to operate systematically search for millilensing events in the entire Fermi sample, because 1) after a decade of operation, thousands of GRBs were detected by the Fermi satellite, new candidates may be found; 2) Fermi/GBM is almost full-time monitoring the whole sky, the millilensing event rate for Fermi GRBs could provide powerful probes of the small-scale structure of the Universe. 

Most recently, \cite{Kalantari2021} applied auto-correlation method to the Fermi sample and find one more millilensing candidate (GRB 090717A). But \cite{Mukherjee2021grb090717} later argued that the light curves of two pulses in GRB 090717 differ at about 5$\sigma$ confidence level, therefore GRB 090717 does not present a compelling example of gravitational lensing. Nevertheless, in the analysis of \cite{Kalantari2021}, spectral properties for each source have not been taken into account (not even the hardness test). In existing case studies, it can be seen that spectral analysis plays a vital role in identifying the authenticity of candidates. For instance, \cite{Mukherjee2021grb950803} found cumulative hardness discrepancies between the two pulses in GRB 950830, thus argued that the case for GRB 950830 involving a gravitational lens may well be considered intriguing – but should not be considered proven. 

In the study of GRB 210812A, \cite{veres2021fermigbm} shows that time-resolved spectrum is a more powerful tool than hardness to help justify the lensing effect. Taking advantage of its two main instruments (GBM and LAT), it is possible to study the $\rm \gamma$-ray spectra of Fermi GRBs in unprecedented detail. In this work, we intend to conduct a comprehensive analysis of the most complete Fermi samples at present to systematically search for millilensing events. We will firstly apply the auto-correlation method to analyze light curves of each source in the sample, and make the preliminary screening of candidates in combination with hardness test (see Section 2). Then we will analyze the time-resolved spectrum for each candidate for further justification (see Section 3). Conclusion and discussion are presented at the end (see Section 4).

\section{Preliminary Candidate Selection} \label{sec:Selection}

\subsection{Data}\label{subsec:Data}

In this work, we use the data observed by the Fermi GBM, which consists of 14 detector modules: 12 Sodium Iodide (NaI) detectors, covering the energies  8 keV - 1 MeV, and two Bismuth Germanate (BGO) detectors, covering 200 keV to 40 MeV \citep{Meegan2009}. The data was downloaded from \textit{Fermi} Science Support Center (FSSC) 's FTP site\footnote{\url{https://heasarc.gsfc.nasa.gov/FTP/fermi/data/gbm/bursts/}}, which contains 3000 GRBs up to April 2021. For each burst, we use the Time-Tagged Events (TTE) data for both spectral and temporal analysis, which records each photon's arrival time with 2 $\mu$s temporal resolution, as well as information regarding in which of the 128 energy channels the photon registered. The prebinned, eight energy channel (CTIME) data was used to carry out the background subtraction. In the temporal analysis (including the auto-correlation test and hardness test), we only use the data from the one NaI with the highest signal to noise ratio. In the spectral analysis, data from BGO detectors are involved. 

\subsection{Auto-correlation test} \label{subsec:autocorr}

For each Fermi GRB, we produce its light curve from the TTE data with 0.01s resolution for short GRB and 0.1s resolution for long GRB. The background was determined by fitting first-order polynomials without the signal part. In order to search for two emission episodes with similar light-curve patterns, we apply the auto-correlation test to each GRB, with the following function:
\begin{equation} \label{auto}
a(\Delta t)=\sum_{t}\frac{f(t)*f(t+\Delta t)}{N\sigma ^2} 
\end{equation}
where $\sigma$ is the standard deviation for the light curve $f(t)$, $\Delta t$ is relative displacement for auto-correlation, and $N$ is the bin number of the light curve.

For each burst, we can plot its $a(\Delta t)-\Delta t$ correlation curve. In principle, if there is no similar emission episodes in the light curve, the $a(\Delta t)-\Delta t$ curve should show a smoothly downward trend with the highest peak at $\Delta t=0$. Otherwise, if similar emission episodes indeed exist, multiple peaks would be superposed on the background decay component. Note that the residual noise may also lead to small wiggles in the $a(\Delta t)-\Delta t$ curve for dim GRBs, here we regard it as an effective peak only when the peak height is 3-$\sigma$ higher than the background. 170 candidates, whose $a(\Delta t)-\Delta t$ curve contains at least one effective peak (peak at $\Delta t=0$ doesn't count), are selected as the preliminary candidates. $\Delta t$ of the highest effective peak could be taken as the possible millilensing time delay, whose uncertainty could be estimated by adding Poisson noise to the original lightcurve \citep[][for a similar approach]{Ukwatta2010,Hakkila2018,Ji2018,veres2021fermigbm}. 

\subsection{Hardness test} \label{subsec:HR}

Hardness test has been widely used to justify the lensing effect \citep{Paynter2021,Kalantari2021,Wang2021,veres2021fermigbm}, based on the hypothesis that the flux ratio between gravitationally lensed pulses should not depend on energy \citep{Paczynski1987}. In this work, we adopt three energy channels to carry out the hardness test: Low energy channel (8-50 keV), Medium energy channel (50-110 keV) and High energy channel (110-323 keV). 

For each preliminary candidate, we first plot their light curves in different energy channels. Based on the auto-correlation results, we can divide each light curve into two similar episodes (we first selected the interval containing the first episode by visually identifying contiguous temporal bins with significant signal, and then selected the second interval with the same length as the first one, but with a certain time delay). Then we define hardness ratios HR$_{\rm HM}$ and HR$_{\rm ML}$ for each episode as

\begin{equation} \label{HR}
{\rm HR}_{ij}=\frac{N_{i}-B_{i}}{N_{j}-B_{j}},
\end{equation}
where $N$ is the total photon counts of the episode, $B$ is the corresponding background photon counts, and $i,j=\left\{L, M ,H\right\}$ indicate the channel index.
The uncertainty of the HR$_{ij}$ could be estimated as 
\begin{eqnarray} \label{HR_error}
\Delta{\rm HR}_{ij}=\left[\frac{N_{i}}{(N_{j}-B_{j})^{2}}+\frac{B_{i}}{(N_{j}-B_{j})^{2}}+\right. \\
\left. \nonumber\frac{N_{j}(N{i}-B_{i})^{2}}{(N_{j}-B_{j})^{4}}+\frac{B_{j}(N{i}-B_{i})^{2}}{(N_{j}-B_{j})^{4}} \right] ^{\frac{1}{2}}
\end{eqnarray}
where Poisson noise is considered. Here, we require the preliminary candidate to pass the Hardness test only when their HR$_{\rm HM}$ and HR$_{\rm ML}$ for different episodes being consistent with the mean value within 1-$\sigma$ region. Eventually we have 4 candidates passing the Hardness test, i.e., GRB 081122A, GRB 081126A, GRB 090717A and GRB 110517B.

\section{Final Candidate Selection} \label{subsec:spectral}

In order to make a better justification on the final candidates, for each candidate, we have analyzed and compared the time integral spectrum and time-resolved spectrum of the two emission episodes in detail. The spectral fitting is performed by using Markov Chain Monte Carlo (MCMC) method with an automatic code “McSpecfit” \citep{Zhang2018}, and in the analysis of time-resolved spectrum, the time bin for each slices are manually selected to ensure sufficient SNR, so as to give a reliable fitting. For each fitting, we adopt four different spectral models, including signal power-law, cutoff power-law, Band function and Blackbody function.  We use the Bayesian Information Criteria (BIC; \cite{1978AnSta...6..461S}) to test the goodness of each model, which shows that in most slices of all candidates, Band function and cutoff power-law models are clearly better than signal power-law and Blackbody models, while cutoff power-law performs slightly better than Band function. Therefore, we decided to use the best fitting parameters of cutoff power-law model to compare the similarity of the two emission episodes. We discuss each candidate in detail below.

\begin{deluxetable*}{cccccccc}\label{table}
\tablecaption{Millilensing GRB candidates in Fermi GBM sample} 
\tablenum{1}
\tablehead{\colhead{GRB name} & \colhead{$\Delta t$}  & \colhead{HR$_{\rm HM}$}  & \colhead{HR$_{\rm HM}$} & \colhead{HR$_{\rm ML}$} & \colhead{HR$_{\rm ML}$}& \colhead{f} & \colhead{M$_{\rm L}$(1+z$_{\rm L}$)} \\ 
\colhead{} & \colhead{(s)}  & \colhead{episode 1} & \colhead{episode 2} & \colhead{episode 1} & \colhead{episode 2} & \colhead{} & \colhead{(M$_\odot$)} }
\startdata
GRB 081126A & 30.6$\pm$0.3 & 0.98$\pm$0.10 & 0.84$\pm$0.11 & 0.59$\pm$0.06 & 0.57$\pm$0.07 & 1.35$\pm$0.02 & $5.1_{-0.3}^{+0.3}\times10^6$ \\
GRB 090717A & 42.1$\pm$0.2 & 0.63$\pm$0.04 & 0.61$\pm$0.07 & 0.44$\pm$0.02 & 0.44$\pm$0.04 & 1.69$\pm$0.01 & $4.02_{-0.05}^{+0.05}\times10^6$ \\
GRB 081122A & 13.7$\pm$0.3 & 0.77$\pm$0.06 & 0.63$\pm$0.11 & 0.67$\pm$0.05 & 0.56$\pm$0.07 & 2.22$\pm$0.07 & $8.6_{-0.4}^{+0.4}\times10^5$\\
GRB 110517B & 17.2$\pm$0.1 & 0.51$\pm$0.06 & 0.54$\pm$0.07 & 0.52$\pm$0.05 & 0.48$\pm$0.04 & 1.04$\pm$0.02 & $2.2_{-1.1}^{+1.1}\times10^7$\\
\enddata
\end{deluxetable*}

\subsection*{GRB 081126A} \label{subsec:GRB 081126A}

\begin{figure*}[!htp]
\renewcommand\thefigure{1}
\centering
\includegraphics[width = 18cm]{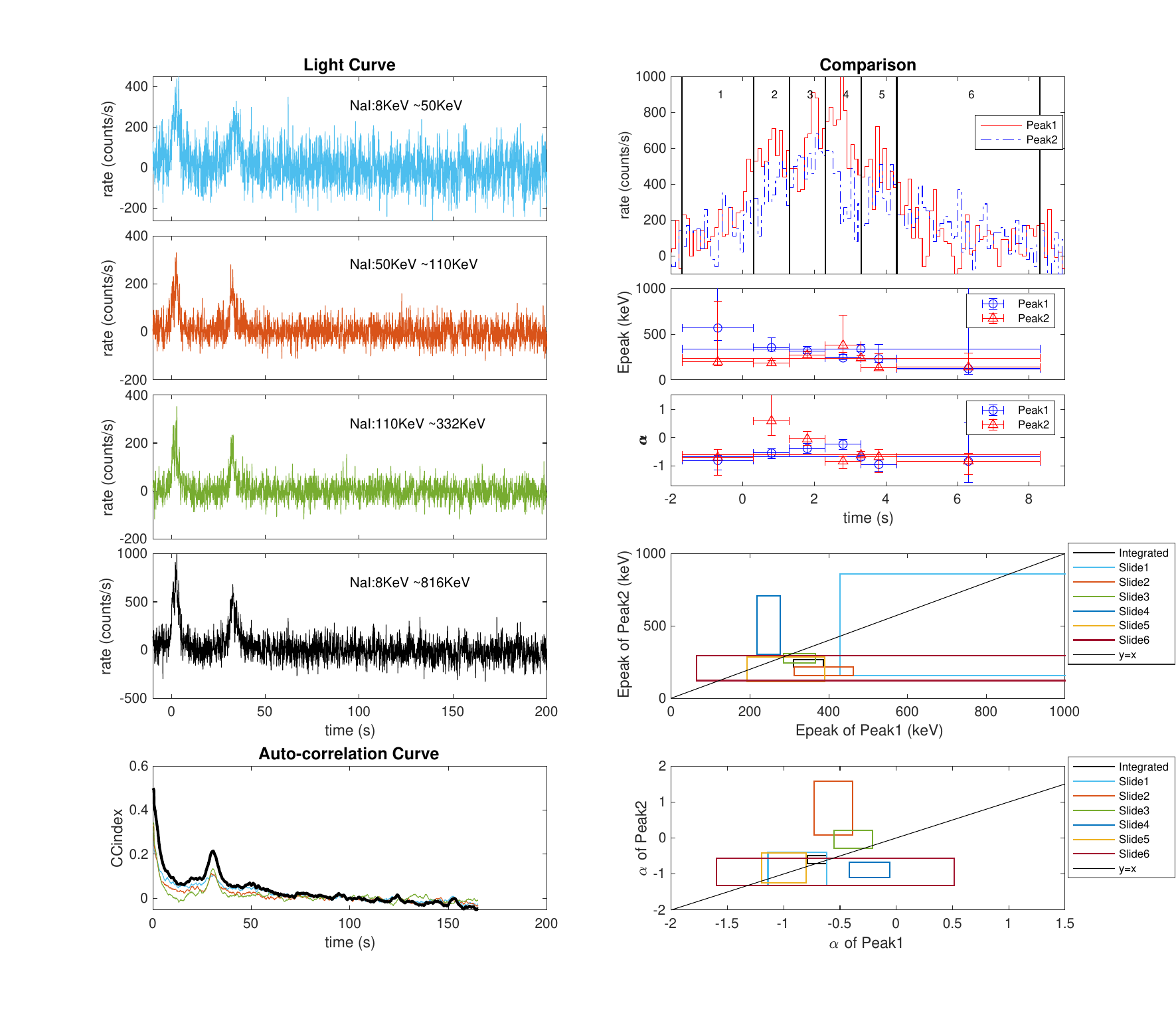}
\caption{Temporal and spectral analysis for GRB 081126A. The left panels show light curves in different energy bands and the corresponding auto-correlation curve for each band. The right panels show comparison of the temporal structure and spectrum parameters of the two emission episodes.
\label{fig:GRB081126A_lightcurve}}
\end{figure*}

GRB 081126A was triggered on 2008-11-26 21:34:09.065 UT ($T_{0}$) by Fermi-GBM (\cite{2008GCN..8559....1H}; trigger number 081126899) with a duration of $54.145\pm0.923$s ($T_{90}$, the time interval between $5\%$ and $95\%$ of the cumulative flux). The analysis results are shown in Figure \ref{fig:GRB081126A_lightcurve}. Auto-correlation analysis suggests the time delay between two emission episodes is $30.6\pm0.3$. We select data from the $T_{0} - 1.7s$ to $T_{0} + 8.3s$ as the first episode and shift it to the second episode with $\Delta t=30.6$ s. The hardness ratios of two episodes are consistent both for the first episode ($\rm HR_{HM}=0.98\pm0.10$, $\rm HR_{ML}=0.59\pm0.06$) and for the second episode ($\rm HR_{HM}=0.84\pm0.11$, $\rm HR_{ML}=0.57\pm0.07$).  Here we performed a so-called ``$\chi^{2}$ test" for testing the light curve similarity of two episodes, which considers the binned lightcurves of the two pulses as representing two distributions and asks if they are consistent with coming from the same parent distribution  \cite{Mukherjee2021a}. Following the same algorithm as \cite{Mukherjee2021a}, we calculate the minimum $\chi^{2}$ value for GRB 081126A, which is $100.94$ for 100 degrees of freedom with the 0.1s time bin, corresponding to a p-value of 0.46.

The time-integrated spectrum parameters of two episodes are $E_{\rm peak1}=336^{+49}_{-26}$, $\alpha_{\rm peak1}=-0.69^{+0.06}_{-0.10}$ and $E_{\rm peak2}=238^{+28}_{-21}$, $\alpha_{\rm peak2}=-0.61^{+0.12}_{-0.10}$. As shown in Figure \ref{fig:GRB081126A_lightcurve}, we divide each episode into 6 slides to compare the time-resolved spectrum. In most slides (4/6), the $\alpha$ and $E_{\rm peak}$ uncertainties of two episodes both overlap within 1-$\sigma$ region. At the same time, the spectrum parameters of two episodes are consistent within 2-$\sigma$ confidence level in every slide for time-resolved spectrum. Considering that 
the detection of GRB 081126A has a relatively high SNR, and based on its excellent performance in light curve and spectrum analysis, we rank it as the first class candidate for millilensing event. Moreover, based on the early detection on the afterglow of GRB 081126A in the UV filters, it is suggested that GRB 081126A should have a redshift of approximately $2.8<z<3.8$ \citep{2008GCN..8564....1H,2008GCN..8589....1B}, which further increases the possibility for GRB 081126A being a millilensing event. 

For a lensed GRB, one can estimate the redshift mass of the lens source with
\begin{equation} \label{Time delay}
M_{L}(1+z_{L})=\frac{c^{3}\Delta t}{2G(\frac{f-1}{\sqrt{f}}+\ln f)}
\end{equation}
where $f$ is the amplification ratio between two images, $z_{L}$ is redshift of the lens source, $G$ is the gravitational constant, and $c$ is the speed of light \citep{Mao1992}. Here $f$ could be calculated with the total counts and background counts of two episodes as $f=(N_{1}-B_{1})/(N_{2}-B_{2})$. For GRB 081126A, we have $f=1.35\pm0.02$. In this case, the redshift mass of the lens source can be estimated as $M_{L}(1+z_{L})=5.1^{+0.3}_{-0.3}\times10^6 M_{\odot}$.

\subsection*{GRB 090717A} \label{subsec:GRB 090717A}

\begin{figure*}[!htp]
\renewcommand\thefigure{2}
\centering
\includegraphics[width = 18cm]{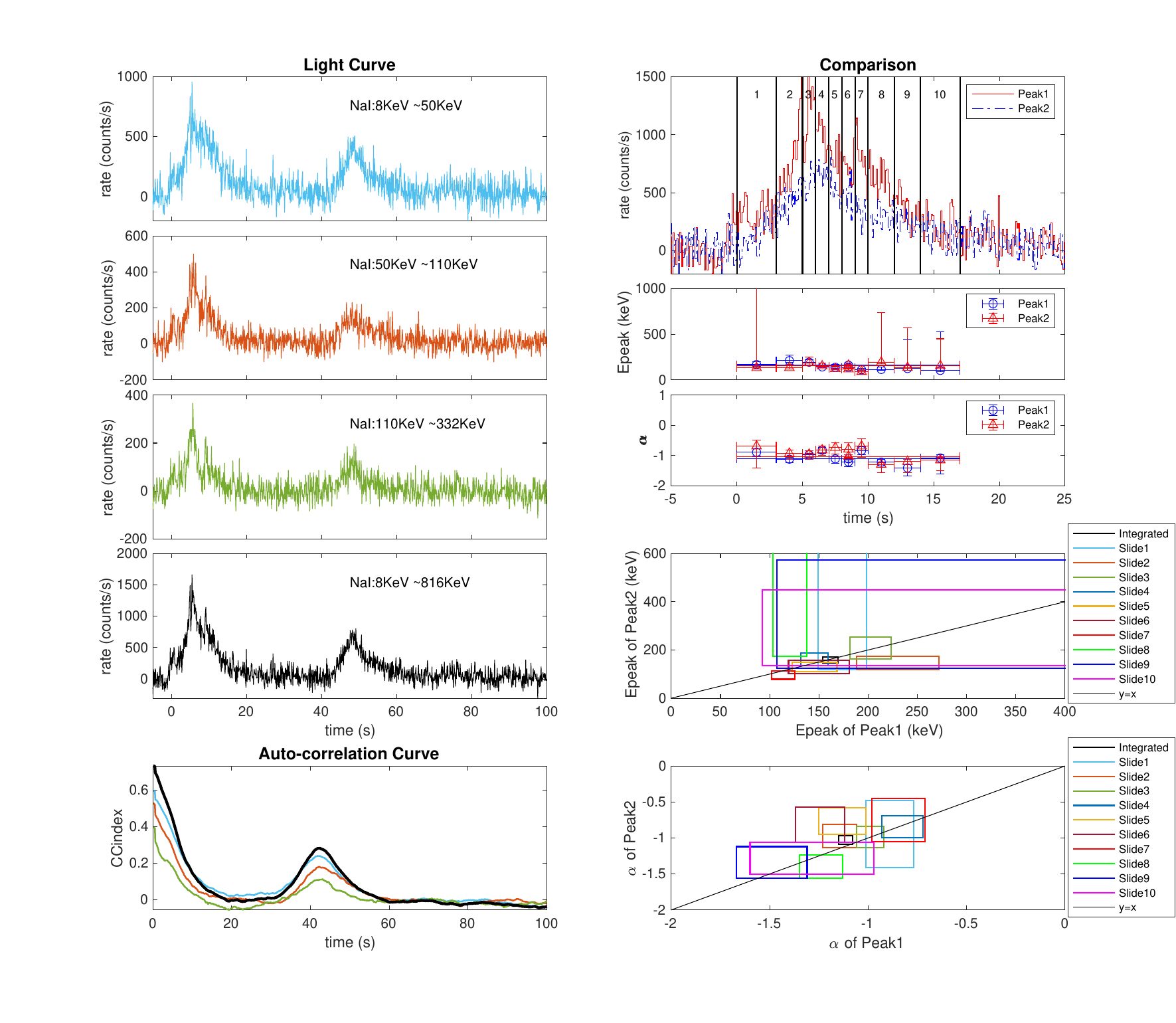}
\caption{Temporal and spectral analysis for GRB 090717A.
\label{fig:GRB090717A_lightcurve}}
\end{figure*}

GRB 090717A was triggered on 2009-07-17 00:49:32.108 UT ($T_{0}$) by Fermi-GBM (\cite{2009GCN..9692....1K}; trigger number 090717034) with a duration of $65.537\pm1.577$s. The analysis results are shown in Figure \ref{fig:GRB090717A_lightcurve}. Auto-correlation analysis suggests the time delay between two emission episodes is $42.1\pm0.2$s. We select data from the $T_{0}$ to $T_{0} + 17s$ as the first episode and shift it to the second episode with $\Delta t=42.1$s. The hardness ratios of two episodes are consistent both for the first episode ($\rm HR_{\rm HM}=0.63\pm0.04$, $\rm HR_{\rm ML}=0.44\pm0.02$) and for the second episode ($\rm HR_{\rm HM}=0.61\pm0.07$, $\rm HR_{\rm ML}=0.44\pm0.04$). The $\chi^{2}$ test obtains $\chi^{2}_{min}=196.7$ for 230 degrees of freedom with the 0.1s time bin, which corresponds to a p-value of 0.95 \footnote{It is worth noticing that the result of ``$\chi^{2}$ test" is very sensitive to the time bin sizes of the light curve. For instance, \cite{Mukherjee2021a} found a relatively low p-value when the time bin for GRB 090717A is adopted as 1.024 s. Comprehensive analysis, especially the time-resolved spectral analysis, is thus essential for justifying millilensing effects.}.

The time-integrated spectrum parameters of two episodes are $E_{\rm peak1}=161^{+8}_{-7}$, $\alpha_{\rm peak1}=-1.11^{+0.03}_{-0.04}$ and $E_{\rm peak2}=158^{+13}_{-12}$, $\alpha_{\rm peak2}=-1.03^{+0.06}_{-0.06}$. As shown in Figure \ref{fig:GRB090717A_lightcurve}, for time-integrated spectrum, the $\alpha$ and $E_{\rm peak}$ of two episodes are remarkably consistent within 1-$\sigma$ confidence level. We then divide each episode into 10 slides to compare the time-resolved spectrum, and the spectrum parameters of two episodes are consistent within 1-$\sigma$ confidence level in every slide except the $\alpha$ of slide 5, 6 and $E_{\rm peak}$ of slide 2, 8, which are consistent within 2-$\sigma$ confidence level. Considering its temporal and spectral analysis results, we rank GRB 090717A as the first class candidate that may have experienced millilensing effect, with the amplification ratio as $f=1.69\pm0.01$, and the redshift mass of the lens source as $M_{L}(1+z_{L})=4.03^{+0.05}_{-0.05}\times10^6 M_{\odot}$.

\subsection*{GRB 081122A} \label{subsec:GRB 081122A}

\begin{figure*}[!htp]
\renewcommand\thefigure{3}
\centering
\includegraphics[width = 18cm]{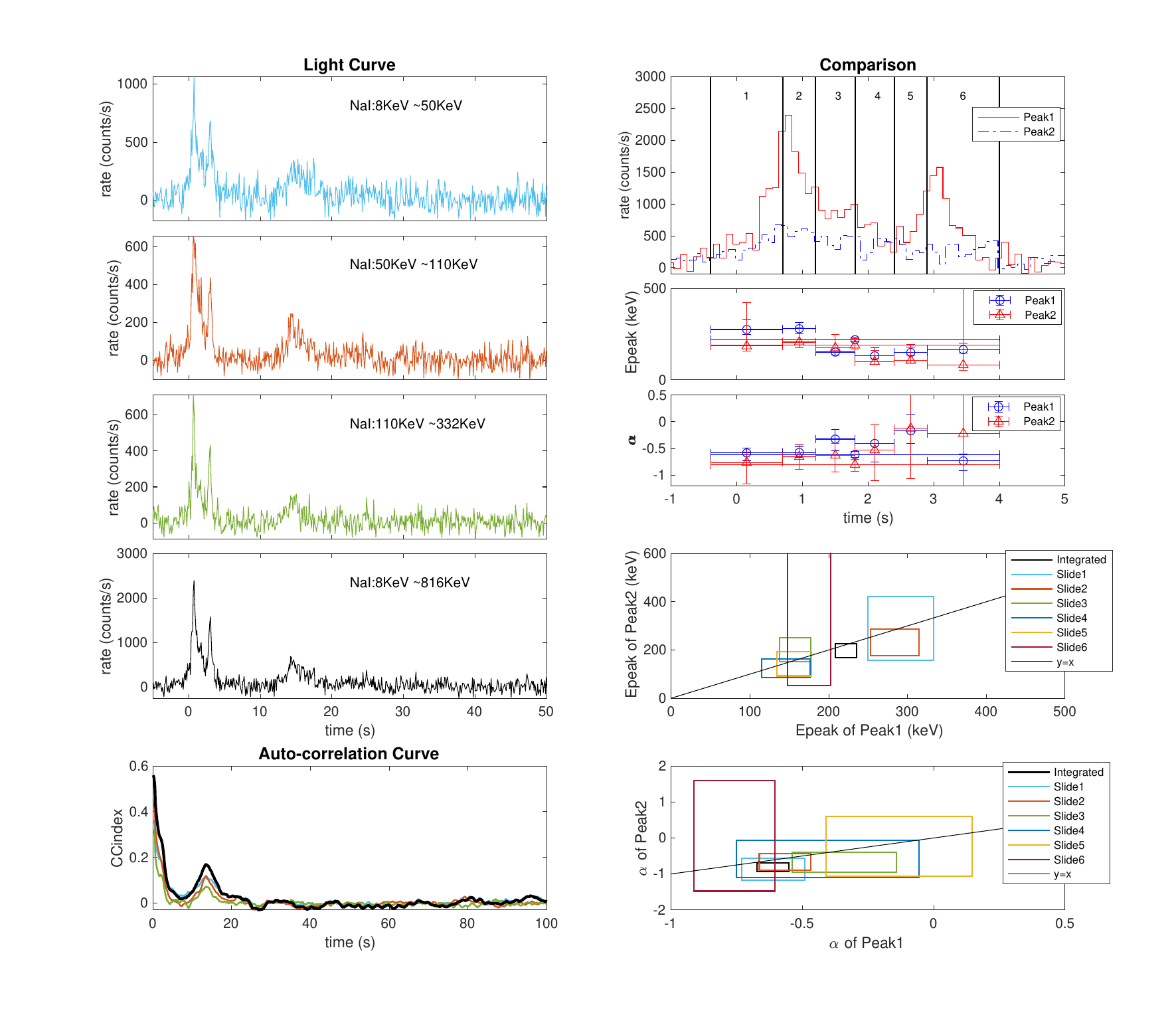}
\caption{Temporal and spectral analysis for GRB 081122A. 
\label{fig:GRB081122A_lightcurve}}
\end{figure*}

GRB 081122A was triggered on 2008-11-22 12:28:12.211 UT ($T_{0}$) by Fermi-GBM (\cite{2008GCN..8549....1M}; trigger number 081122520) with a duration of $23.30\pm2.11$s. The analysis results are shown in Figure \ref{fig:GRB081122A_lightcurve}. Auto-correlation analysis suggests the time delay between two emission episodes is $13.7\pm0.3$s. We select data from the $T_{0} - 0.4s$ to $T_{0} + 4.0s$ as the first episode and shift it to the second episode with $\Delta t=13.7$s. The hardness ratios of two episodes are consistent both for the first episode ($\rm HR_{\rm HM}=0.77\pm0.06$, $\rm HR_{\rm ML}=0.67\pm0.05$) and for the second episode ($\rm HR_{\rm HM}=0.63\pm0.11$, $\rm HR_{\rm ML}=0.56\pm0.07$). The $\chi^{2}$ test obtains $\chi^{2}_{min}=97.98$ for 88 degrees of freedom with the 0.05s time bin, which corresponds to a p-value of 0.21. Here 0.05s time bin is adopted because the total duration of GRB 081122A is relatively short.

The time-integrated spectrum parameters of two episodes are $E_{\rm peak1}=220^{+16}_{-11}$, $\alpha_{\rm peak1}=-0.61^{+0.06}_{-0.05}$ and $E_{\rm peak2}=189^{+37}_{-20}$, $\alpha_{\rm peak2}=-0.80^{+0.10}_{-0.12}$.
As shown in Figure \ref{fig:GRB081122A_lightcurve}, we divide each episode into 6 slides to compare the time-resolved spectrum. The spectrum parameters of two episodes are consistent within 1-$\sigma$ confidence level for each slide. The overall spectral analysis results provide positive evidence to support GRB 081122A to be a millilensing candidate. However, we notice that there are two peaks in the first emission episode but the light curve of the second episode seems no such a feature which may be due to the low SNR. We thus rank it as the second class candidate that may have experienced millilensing effect. With the the amplification ratio $f=2.22\pm0.07$, the redshift mass of the lensing source can be estimated as $M_{ L}(1+z_{L})=8.6_{-0.4}^{+0.4}\times10^5 M_{\odot}$.

\subsection*{GRB 110517B} \label{subsec:GRB 110517B}

\begin{figure*}[!htp]
\renewcommand\thefigure{4}
\centering
\includegraphics[width = 18cm]{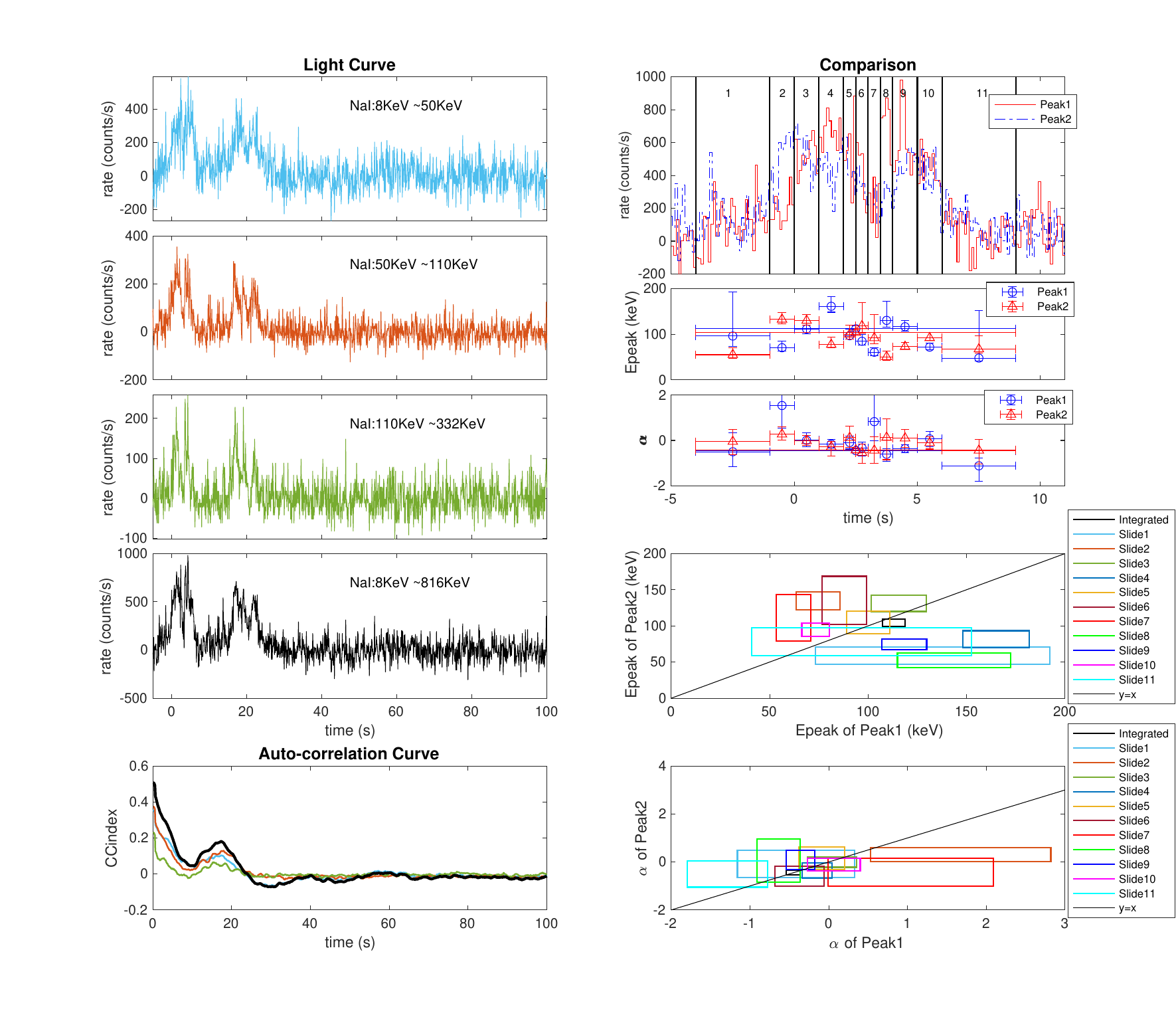}
\caption{Temporal and spectral analysis for GRB 110517B. 
\label{fig:GRB110517B_lightcurve}}
\end{figure*}

GRB 110517B was triggered on 2011-05-17 13:44:47.600 UT ($T_{0}$) by Fermi-GBM (trigger number 110517573) with a duration of $23.04\pm0.36$s. The analysis results are shown in Figure \ref{fig:GRB110517B_lightcurve}. Auto-correlation analysis suggests the time delay between two emission episodes is $17.2\pm0.1$s. We select data from the $T_{0} - 4s$ to $T_{0} + 9s$ as the first episode and shift it to the second episode with $\Delta t=17.2$s. The hardness ratios of two episodes are consistent both for the first episode ($\rm HR_{\rm HM}=0.51\pm0.06$, $\rm HR_{\rm ML}=0.52\pm0.05$) and for the second episode ($\rm HR_{\rm HM}=0.54\pm0.07$, $\rm HR_{\rm ML}=0.48\pm0.04$). The $\chi^{2}$ test obtains $\chi^{2}_{min}=185.4$ for 130 degrees of freedom with the 0.1s time bin, which corresponds to a p-value of 0.001.

The time-integrated spectrum parameters of two episodes are $E_{\rm peak1}=112^{+7}_{-5}$, $\alpha_{\rm peak1}=-0.45^{+0.09}_{-0.09}$ and $E_{\rm peak2}=103^{+6}_{-5}$, $\alpha_{\rm peak2}=-0.43^{+0.10}_{-0.10}$. As shown in Figure \ref{fig:GRB110517B_lightcurve}, for time-integrated spectrum, the spectrum parameters of two episodes are consistent within 1-$\sigma$ confidence level. After we divide each episode into 11 slides to compare the time-resolved spectrum, the $\alpha$ of two episodes are consistent within 1-$\sigma$ region in every slide, but the $E_{\rm peak}$ are only consistent within 1-$\sigma$ region in 3 slides ($27.3\%$), and the spectrum evolution of the $E_{\rm peak}$ show different trend in two episodes , which cuts down the possibility for millilensing event. We notice that there are multiple peaks in both emission episodes, where the width distribution of these peaks is similar, but the peak amplitude distribution exists some differences in detail (that is why we get a small p-value in the $\chi^{2}$ test). Considering its temporal and spectral analysis results, we rank GRB 110517B as the second class candidate for millilensing event. With the amplification ratio $f=1.04\pm0.02$, the redshift mass of the lensing source can be estimated as $M_{L}(1+z_{L})=2.2_{-1.1}^{+1.1}\times10^7 M_{\odot}$.

\section{Conclusion and discussion} \label{sec:summary}

It has long been proposed that GRBs have potential to be gravitationally lensed into multiple images, due to their high redshift nature. When the lens source has mass $10^4~\sim~10^7  M_{\odot}$ (such as intermediate-mass black hole, compact dark matter, star cluster or population III star), the delay time between different images would be in order of $10^{-1}\sim10^{2}$ s, which is comparable with the typical duration of GRBs. In this case, signals from two images would be collected within one single trigger, manifesting as one GRB that contains two (or even more) emission episodes with similar properties in both temporal and spectral domain. Such kind of events are usually called millilensing events. 

In previous works, a few GRBs have been proposed as the millilensing candidates, i.e., BATSE GRB 950830, Fermi GRB 200716C \footnote{Note that this burst is also included in the 3000 GRBs we searched. Since this source has been discussed in detail in two independent works, we do not list it as the candidate we found.} and Fermi GRB 210812A. In this work, we conduct a systemic search for millilensing events from the Fermi GRB sample (3000 GRBs up to April 2021). We select preliminary candidates by performing auto-correlation test and hardness test, and then we analyze and compare the time integrated spectrum and time-resolved spectrum of the similar emission episodes. Eventually we find 4 interesting candidates which may have experienced millilensing effect, e.g. GRB 081126A, GRB 090717A, GRB 081122A and GRB 110517B. We rank GRB 081126A and GRB 090717A as the first class candidate based on their excellent performance both in temporal and spectrum analysis. We rank GRB 081122A and GRB 110517B as the second class candidates (suspected candidates), mainly because their two emission episodes show clear deviations in part of the time-resolved spectrum or in the time-integrated spectrum. The ratio of the number of lensed candidates to the total number of GRB in our sample is $2/3000 \sim 4/3000$. The redshift mass of the lensing source for our selected candidates are within the range of $8.6\times10^5 M_{\odot}<M_{L}(1+z_{L})<2.2\times10^7 M_{\odot}$, which are very likely star clusters or super-massive black holes.

For a given GRB, the lensing optical depth can be written as \cite[][for details]{Zhou2021a,Zhou2021b}

\begin{eqnarray}\label{eq1}
\nonumber \tau(f_{L},z_{S})=\int_0^{z_{S}}d\chi(z_{L})(1+z_{L})^2n_{L}(f_{L})\sigma(M_{\rm Len},z_{L},z_{S}) \\ 
=\frac{3}{2}f_{L}\Omega_{m}\int_0^{z_{S}}dz_{L}\frac{H_0^2}{cH(z_{L})}
\frac{D_{L}D_{\rm LS}}{D_{S}}(1+z_{L})^2y^2_{\max}(R_{\rm f,max})
\end{eqnarray}
where monochromatic mass distribution for lens source is assumed, $f_{L}$ represents the fraction of the mass of lens with mass $M_{L}$ to the total matter mass (dark matter mass + baryon mass), $z_{S}$ is the redshift of the GRB, $z_{L}$ is the redshift of the lens, $n_{L}$ is the comoving number density of the lens, $H(z_{L})$ is the Hubble parameter at $z_{L}$, $H_0$ is the Hubble constant, and $\Omega_{m}$ is the present density parameter of matter. Here we define the  the maximum value of normalized impact parameter as $y_{\max}(R_{\rm f,max})=R_{\rm f,max}^{1/4}-R_{\rm f,max}^{-1/4}$, by requiring that the flux ratio of two lensed images is smaller than a critical value $R_{\rm f,max}$ (in this work we take $R_{\rm f,max}=5$). If we accumulate a considerable number of GRBs with redshifts satisfying $N(z_{S})$ distribution, the integrated optical depth of all these GRBs would be
\begin{equation}\label{eq2}
\bar{\tau}(f_{L})=\int dz_{S}\tau(f_{L},z_{S})N(z_{S}).
\end{equation}
Consequently, the expected lensing fraction for GRBs would be
\begin{equation}\label{eq3}
{\cal F}=\frac{N_{\rm Lensed~GRB}}{N_{\rm GRB}}=(1-e^{-\bar{\tau}(f_{L})}).
\end{equation}

In this work, our results suggest that ${\cal F}=2/3000 \sim 4/3000$ (assuming $M_{\rm L}\sim10^{6} M_{\odot}$), inferring $f_{\rm L}=(0.5\sim1)\times10^{-2}$. It is worth noting that although we have made a detailed analysis of each source in the sample, it can not be ruled out that there are still some possible candidates who have not been screened out. The main reason is that when the time delay is shorter than the intrinsic time scale of the burst, the signals of different images will overlap. This situation is not easy to be distinguished through auto-correlation analysis, especially when the signal itself has a complex structure. To be conservative, here we suggest to use $0.5\times10^{-2}$ as a lower bound estimate for the fraction of the mass of lens objects with mass $M_{\rm L}\sim10^{6} M_{\odot}$ to the total matter mass, inferring that the density parameter of lens objects with mass $M_{\rm L}\sim10^{6} M_{\odot}$ is larger than $1.5\times10^{-3}$. 

Up to now, all candidates (including our findings and previous work findings) are proposed solely based on the gamma-ray data analysis. However, the physical origin of gamma-ray burst (GRB) prompt emission is still poorly understood \citep{Zhang2018book}. It cannot be excluded that a GRB might have two emission episodes with inherently similar temporal and spectral characteristics \citep{Lan2018}. Multi-band observations would be essential to finally determine whether a GRB has really experienced the gravitational lensing effect \citep{Chen2021}. With the successful operation of many sky survey projects in multiple bands, such as all-sky gamma-ray monitors (e.g. Gravitational wave high-energy Electromagnetic Counterpart All-sky Monitor; \citep{GECAM}), sky survey detectors in the X-ray band (e.g. Einstein Probe; \citep{Yuan2018}), and wide field of view monitoring system in optical band (e.g. Ground-based Wide Angle Camera system; \citep{SVOM}), more lensed GRBs are expected to be detected and accurately certified in the future.

\acknowledgments
We thank Peter Veres and Hou-Jun L\"u for helpful discussions and the anonymous referee for the helpful comments that have helped us to improve the presentation of the paper. HG is supported by the National Natural Science Foundation of China (NSFC) under Grant No. 12021003. B.B.Z acknowledges support by the National Key Research and Development Programs of China (2018YFA0404204), the National Natural Science Foundation of China (Grant Nos. 11833003, U2038105), the science research grants from the China Manned Space Project with NO.CMS-CSST-2021-B11, and the Program for Innovative Talents, Entrepreneur in Jiangsu.

\bibliography{Searching_gravitational_lensing_echoes_of_gamma-ray_bursts_in_the_data_of_Fermi_GBM.bib}{}
\bibliographystyle{aasjournal}

\end{document}